\documentclass[10pt, conference]{IEEEtran}
\usepackage[left=0.625in,right=0.625in,top=0.75in,bottom=0.99 in]{geometry}
\IEEEoverridecommandlockouts
\usepackage{cite}
\usepackage{amsmath}
\usepackage{amssymb}
\usepackage{amsfonts}
\usepackage{graphicx}
\usepackage{color}
\usepackage{subfig}
\usepackage{stackengine}
\usepackage{comment}
\usepackage[normalem]{ulem}
\usepackage{tabularx}
\usepackage{bm}
\usepackage{dsfont}
\usepackage{mathtools}
\usepackage{lettrine}
\usepackage[font=scriptsize]{caption}

\usepackage[hidelinks]{hyperref} 

\usepackage[flushleft]{threeparttable} 
\usepackage{booktabs}

\usepackage{textcomp}
\usepackage{xcolor}

\usepackage[linesnumbered, ruled]{algorithm2e}
\SetKwRepeat{Do}{do}{while}%

\SetCommentSty{mycommfont}
\usepackage{algpseudocode}

\def\BibTeX{{\rm B\kern-.05em{\sc i\kern-.025em b}\kern-.08em
    T\kern-.1667em\lower.7ex\hbox{E}\kern-.125emX}}
    
\algnewcommand\algorithmicforeach{\textbf{for each}}
\algdef{S}[FOR]{ForEach}[1]{\algorithmicforeach\ #1\ \algorithmicdo}

\usepackage{orcidlink}

\usepackage{xspace}
\usepackage{ifthen}
\usepackage[inline]{enumitem}

\usepackage{algcompatible}

\newboolean{showcomments}
\setboolean{showcomments}{true}
\ifthenelse{\boolean{showcomments}}
{
}

\usepackage{subfig}
\usepackage{fancyhdr}

\begin{document}

\title{\huge Scalability Assurance in SFC provisioning via Distributed Design for Deep Reinforcement Learning}

\author{
\IEEEauthorblockN{Murat Arda Onsu$^1$, Poonam Lohan$^1$, Burak Kantarci$^1$, Emil Janulewicz$^2$,}\\
\IEEEauthorblockA{\textit{$^1$University of Ottawa, Ottawa, ON, Canada}\\
\textit{$^2$Ciena, 383 Terry Fox Dr,
Kanata, ON K2K 2P5, Canada}\\
$^1$\{monsu022, ppoonam, burak.kantarci\}@uottawa.ca,~$^2$\{ejanulew\}@ciena.com}
\vspace{-0.2in}}

\maketitle
\begin{abstract}
High-quality Service Function Chaining (SFC) provisioning is provided by the timely execution of Virtual Network Functions (VNFs) in a defined sequence. Advanced Deep Reinforcement Learning (DRL) solutions are utilized in many studies to contribute to fast and reliable autonomous SFC provisioning. However, under a large-scale network environment, centralized solutions might struggle to provide efficient outcomes when handling massive demands with stringent End-to-End (E2E) delay constraints. Therefore, in this paper, a novel distributed SFC provisioning framework is proposed, where the network is divided into several clusters. Each cluster has a dedicated local agent with a DRL module to handle the SFC provisioning of demands in that cluster. Also, there is a general agent that can communicate with local agents to handle the requests beyond their capacity. The DRL module of local agents can be applied under different configurations of clusters independent of different numbers of data centers and logical links in each cluster. Simulation results demonstrate that utilizing the proposed distributed framework offers up to 60\%  improvements in the acceptance ratio of service requests in comparison to the centralized approach while minimizing the E2E delay of accepted requests.
\end{abstract}
\hypersetup{nolinks=true}
\begin{IEEEkeywords} SFC Provisioning, Deep-Reinforcement Learning, Distributed Approach, Scalability,  Priority Points
\end{IEEEkeywords}

\section{Introduction} \label{sec:1}
Network Function Virtualization (NFV) reduces OPEX and CAPEX while increasing agility and flexibility by utilizing virtualization technology, decoupling software from physical devices, and deploying functions on general-purpose hardware, such as data centers (DCs) \cite{new_4}. Service Function Chaining (SFC) provisioning, which sequences Virtual Network Functions (VNFs) to deliver services such as Cloud Gaming (CG), Augmented Reality (AR), VoIP, Video Streaming (VS), Massive IoT (MIoT), and Industrial 4.0 (Ind 4.0), maximizes the advantages of NFV \cite{13}\cite{2}. However, SFC provisioning faces key challenges, such as efficient resource allocation, sequential VNF executions, massive demands, and E2E delay constraints.

Several researchers have proposed Deep Learning (DL)-assisted algorithm and Deep Reinforcement Learning (DRL) algorithms \cite{8}, \cite{new_1} for optimal and automated SFC provisioning due to their better decision-making process and adaptability to varying service demands. DRL \cite{DRL} combines the features of DL which are inspired by artificial neural networks \cite{dl} as a model architecture and RL \cite{RL} for algorithm part to provide robust training in terms of complexity and accuracy. However, existing studies do not address large-scale environments with numerous data centers (DCs) and a substantially high volume of SFC demands. In such networks, a single centralized model cannot handle optimal VNF placements for massive SFC demands. Therefore, a distributed framework 
is proposed for SFC provisioning in this work. This approach divides the centralized task into multiple subtasks for different local agents. Network DCs are segmented into distinct clusters, with each cluster and its incoming SFC requests managed by a local agent. The general agent oversees actions of the local agents, assists them with tasks beyond their scope, and maintains inter-cluster connections.

Local agents operate independently, allowing clusters to differ in the number of DCs and logical links. This requires a DRL model architecture that remains consistent across different cluster configurations; otherwise, each cluster would need a new model and retraining, which is inefficient. To avoid information loss from dimension mismatches, the DRL model must handle the varying dimensions of features in the environment while ensuring fixed-dimension input data. To address this, we use an advanced DRL model architecture with multiple input layers for different features, independent of cluster and network configuration changes, from our previous work \cite{onsu2024unlocking}. 
  The main contributions of this work are summarized as follows:
\begin{itemize}
    \item A scalable distributed design is proposed to handle massive SFC requests in a large-scale network and is compared with the centralized approach. 
    \item  SFC provisioning workload is divided among multiple local agents
    and a general agent handles those requests that can not be satisfied by local agents due to resource constraints improving the overall acceptance ratio. 
    \item Cluster-to-cluster packet transmission is proposed for destinations outside the scope of the local agent. This method, handled by the general agent, reduces path discovery time and minimizes the E2E delay of accepted SFC requests.
\end{itemize}

The paper is organized as follows. Section \ref{sec:2} presents relevant works. Section \ref{sec:3} describes system model and problem formulation. The proposed distributed design and its components are explained in Section \ref{NovelAI}. Section \ref{sec:4} provides numerical results and discussions, and Section \ref{sec:5}  concludes the paper.

\section{Related Work} \label{sec:2}
There are several studies related to SFC provisioning and VNF placement in the literature. In \cite{rel1}, researchers introduce DRL into SFC provisioning to improve cost-efficiency in dynamic environments. However, in \cite{rel1}, the DRL model architecture state does not take overall system information, but it is constrained with SFC information. Much more advanced AI model architectures with an adaptable input state can make performance robust. Another proposed cost-efficient SFC provisioning solution is presented in \cite{rel2}, in which researchers use Mixed Integer Linear Programming (MILP) to formalize the SFC chaining and placement problem in a mathematical model for minimizing the operational cost. In addition to cost, this research also considers the end-to-end delay and acceptance rate of SFC demands. The study in \cite{rel6} reports the limitations of the traditional methods of efficiently embedding SFCs, such as the complexity of the network states, high-speed computational requirements, and enormous service requests, and proposes the DRL method due to its promising way of tackling these problems. Additionally, they propose an enhanced DDPG model to handle slow convergence problems and minimize the end-to-end delays in edge clouds for the efficient SFC embedding problem. However, in \cite{rel2}, VNF diversity and SFC types are not considered, and, such as \cite{rel6}, research is done in a fixed node environment.

Moreover, in \cite{rel3}, reliability-aware multi-domain SFC placement problems are studied through a multi-objective optimization model to minimize resource consumption and placement operating costs, and a DRL-based model is introduced without consideration of the delay in SFC acceptance and the processing time requirement of the massive amount of requests. The DRL-based approach is also used in \cite{rel4} to tackle the NP-hardness of the multi-resource SFC scheduling and minimize the average flow completion time, wherein the offline implementation is extended to online SFC scheduling. The DRL model architecture, in \cite{rel4}, is not adaptable for different network configurations since input state size changes with the number of servers and states. Therefore, if this information is changed, the model architecture will be changed as well, and the model will be required to be trained again.  Research in \cite{rel7} also presented a priority-aware deployment framework for autoscaling and multi-objective SFCs for satisfying various QoS requirements to avoid resource congestion. In \cite{rel7}, although researchers work on the high number of SFC rates, the inference time requirement of the AI model to handle massive amount of SFC requests in a given period is not considered. In our study, we provide a scalable solution for SFC provisioning while considering all the above-stated research gaps.

\begin{figure}
    \centering
    \includegraphics[width=1.\linewidth]{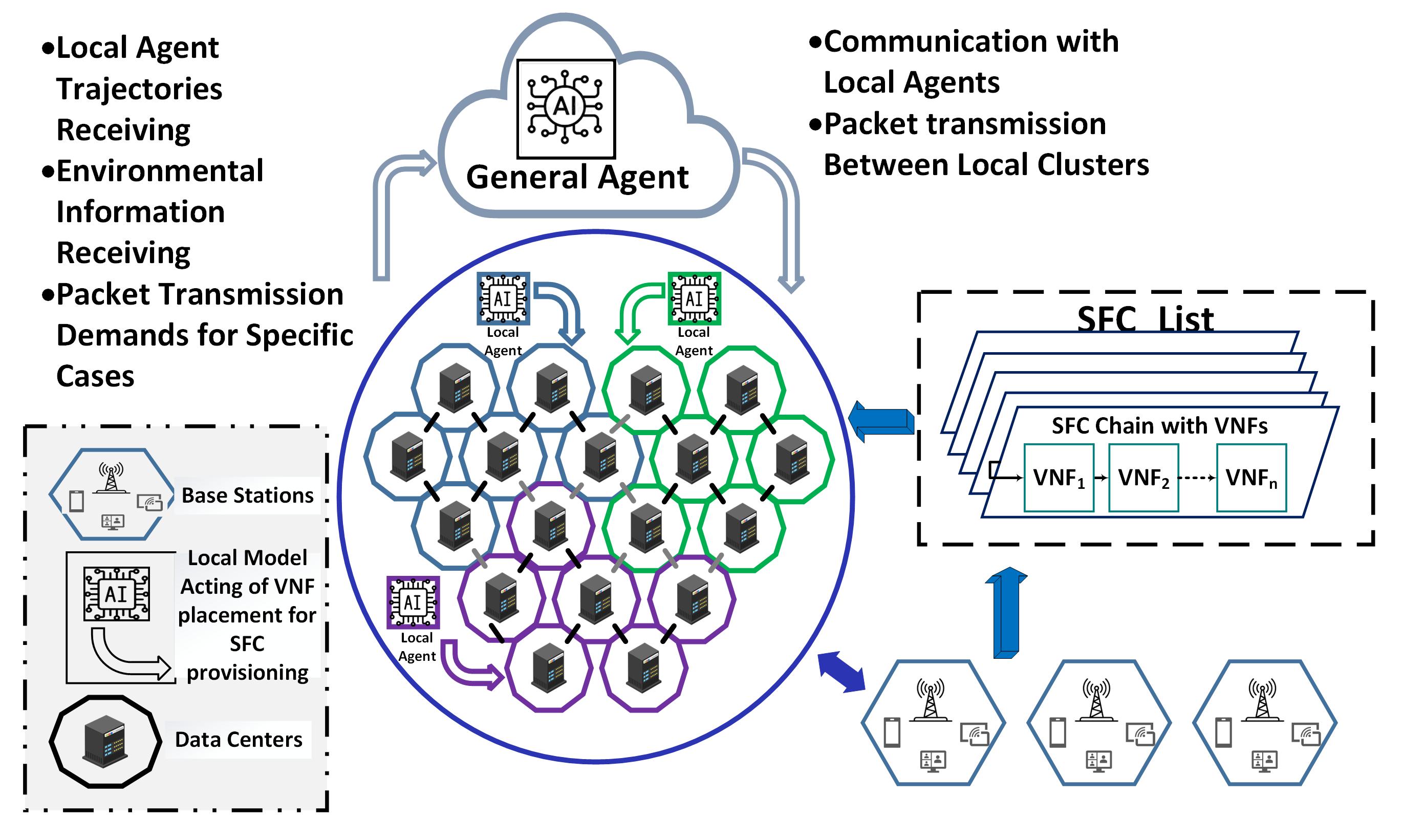}
    \caption{System model example with different Clusters of VNFI-enabled DCs}
    \label{fig: system model}
\end{figure}

\section{System  Model and Problem Formulation} \label{sec:3}

\subsection{System Model and SFC Request Attributes} \label{system}
\figurename \ref{fig: system model} illustrates the network environment, which consists of multiple clusters of VNF instance (VNFI)-enabled data centers (DCs) on which different VNFs can be installed and executed. A specific local agent is assigned to each cluster to handle the SFC requests received by DCs of that cluster. Users’ equipment put their different types of service requests to the network through the base stations connected to DCs as depicted in \figurename \ref{fig: system model}. There is also one global agent that observes the overall systems and helps the local agents to communicate with each other in scenarios where the DCs of the local agent's cluster are out of resources for further VNF placements or the destination of the SFC request is outside the local agent's cluster. A detailed explanation of the  global agent role is provided in Section IV. 

Let the whole network topology be represented by an undirected graph $\mathcal{G(\mathcal{N},\mathcal{L})}$, 
where $\mathcal{N}$ and $\mathcal{L}$ denote the set of VNFI-enabled nodes and logical links among nodes, respectively. Furthermore, The whole network topology is divided into different clusters using a constrained K-means algorithm \cite{constrainedKMEANS} with a limit on the maximum number of nodes in one cluster. Let $C$ be the total number of clusters in the network topology which depends on the network size and maximum number of nodes limit in one cluster, i.e. $\mathcal{G}(\mathcal{N},\mathcal{L})=\{\mathcal{G}_1(\mathcal{N}_1,\mathcal{L}_1),\mathcal{G}_2(\mathcal{N}_2,\mathcal{L}_2),..., \mathcal{G}_C(\mathcal{N}_C,\mathcal{L}_C), \mathcal{L}_x\}$.
Where $\mathcal{G}_c(\mathcal{N}_c, \mathcal{L}_c)$
for $c=\{1,2,..., C\}$ represents the undirected graph with $\mathcal{N}_c$ and $\mathcal{L}_c$ denoting the set of DCs and logical links, respectively, for cluster $c$ ;
and $ \mathcal{L}_x $ denotes the logical links among nodes of different clusters. Let $\mathcal{S}_{i^c}$ denote storage capacity in GB and $\mathcal{Q}_{i^c}$ denote computational capacity (based on CPUs and RAM) in cycles/sec of DC $i^c \in \mathcal{N}_c$. $B_{ij}$ represents the bandwidth capacity in bps (bits per second) of the logical link between DCs $i$ and $j$. Please note that here both intra-cluster and inter-cluster logical links are considered. 

 SFC requests, on the other hand, consist of a VNFs chain, and to satisfy these demands, all VNFs in the chain must be placed in the appropriate DCs and processed in sequential order. The set of SFC service types served by the service provider is denoted by $S$ and the set of all VNFs constituting these SFCs is presented by $V$. $\mathcal{F}^s= (v_1^s\rightarrow v_2^s \rightarrow...v_{N_s}^s)$ represents a total $N_s$ VNFs to be executed in sequence for satisfying the request $s\in S$. Furthermore, all features of each service request $s\in S$ are presented by a tuple
$\{\mathcal{F}^s,   \mathcal{D}^s, \mathcal{B}^s, \lambda_s \}$, where $\mathcal{D}^s$ denotes E2E delay tolerance, $\mathcal{B}^s$ denotes bandwidth capacity needed,  and $\lambda_s$ denotes bundle size of request $s\in S$. It is assumed that multiple same or different VNFI can be placed/installed to each DC and can be allocated to different SFC requests $s\in S$. To process a particular VNF of an SFC request, the same type of VNFI should be installed/placed in DC and allocated to that SFC till its processing time is over. An SFC request can access an installed VNFI when it is free after completing the previously allocated VNF processing. If all installed VNFI are occupied, in that case, a new VNFI can be installed based on the availability of sufficient resources with DCs.   $n_{i^c}^v$  is an integer variable representing the number of VNFI $v\in V$ placed on DC $i^c \in \mathcal{N}_c$. The computational, storage resource requirement, and processing time for VNFI $v \in V$ are denoted by $Q^v$, $S^v$, and $t^v$ respectively. 
Detailed characteristics of SFCs and VNFs are discussed in the numerical section \ref{sec:4}. 
Ensuring efficient placement and allocation of VNFs to meet SFC requests is crucial for improving network performance, as measured by the acceptance ratio while adhering to the E2E delay tolerance limits specified by the SFC requests.

\subsection{Problem Formulation}
 With $A_s^c$ denoting the number of SFC requests served of type $s\in S$ for cluster $c=\{1,2,..., C\}$,
 the total acceptance ratio of the overall network, $A_{r}={\sum_{c=1}^C \sum_{s\in S}\mathcal{A}_s^c}/{\sum_{s\in S} \lambda_s }$, can be defined as the total number of all types of SFC requests served within their E2E delay limit by all clusters divided by the total number of requests generated by all types of request bundles. The total acceptance ratio can be maximized by efficient placements of VNFs on DCs and allocating these VNFs to upcoming SFC requests in each cluster while fulfilling the  resource constraints of DCs, bandwidth capacity constraints of logical links, and E2E delay constraints of SFC requests, that is mathematically expressed as follows:
 \begin{align*}
       &(\mathcal{P}):\;\underset{\mathbf{n},\boldsymbol{\delta}}{\text{maximize}}\; A_r={\sum_{c=1}^C \sum_{s\in S}\mathcal{A}_s^c}/{(\sum_{s\in S} \lambda_s)}\\
 &\text{s.t.:}\; (C1):\sum_{v\in V} n_{i^c}^v Q^v \leq \mathcal{Q}_{i^c}, \; \forall i^c \in \mathcal{N}_c, \; \forall c=\{1,2,..,C\}\\
 & (C2):  \sum_{v\in V} n_{i^c}^v S^v \leq \mathcal{S}_{i^c}, \; \forall i^c \in \mathcal{N}_c, \; \forall c=\{1,2,..,C\}\\
 &(C3): \sum_{{i}\in \mathcal{N}} \delta_{i}^{v_k^s} \mathds{1}(n_{i}^{v_k^s}>0) = 1, \; \forall s \in S, \;\;\forall k=\{1,2,.....N_s\}\\
 & (C4): \sum_{\lambda_s \forall s\in S}\sum_{k=1}^{N_s-1} \delta_{i}^{v_k^s} \delta_{j}^{v_{k+1}^s} \mathcal{B}^s \leq B_{ij}, \; \forall i,j \in \mathcal{N},  \;\; i\neq j  \\
 & (C5): t_{P_g}^s+  t_{P_r}^s \leq \mathcal{D}^s \;; \forall s\in S
    \end{align*}
    
Here, constraints (C1) and (C2), respectively, state that for each DC $i^c$ in every cluster $c$, the computational and storage capacity of the placed VNFIs should not be greater than its total computational and storage capacity. Constraint (C3) ensures that each VNF function in SFC request $s$ should be processed by only one DC; where, $\delta_{i}^{v_k^s}$ is a binary variable which is set equal to $1$ if $k^{th}$ VNF in SFC $s$ is processed at node $i$ with condition $(n_{i}^{v_k^s}>0)$; otherwise, it is set to zero. If sufficient resources are not available within the cluster then the global agent assigns another cluster to fulfill the SFC request. Since VNFs of the same SFC can be processed by DCs of different clusters, the notation $i$ is used to denote a DC that may belong to any cluster of the overall network.  Constraint (C4) is related to logical link bandwidth constraint and states that the bandwidth resource occupied by all SFCs at an instant should not exceed the bandwidth capacity of all intra-cluster ($\mathcal{L}_1,\mathcal{L}_2,.., \mathcal{L}_C)$ and inter-cluster logical links $\mathcal{L}_x$.
Lastly, constraint (C5) confirms that for successful completion of the SFC request the propagation and processing delay of SFC should not be more than the E2E delay tolerance limit of that SFC request. Propagation delay, $ t_{P_g}^s= \sum_{k=1}^{N_s-1}  \sum_{i\in \mathcal{N}}  \sum_{j\in \mathcal{N}} \delta_i^{v_k^s} \delta_j^{v_{k+1}^s} t_{ij}^{P}$, depends upon the distance weight $d_{ij}$ between nodes $i$ and $j$ which serves two consecutive VNFs for SFC request $s$ and the speed of light, $c=3*10^8 km/s$ (considering optical fiber as logical link among DCs) i.e. $t_{ij}^P=d_{ij}/c$. Processing delay, $t_{P_r}^s=\sum_{k=1}^{N_s}  \sum_{i\in \mathcal{N}} \delta_{i}^{v_k^s} (w^{v_k^s}+t^{v_k^s})$, depends upon the waiting time $w^{v_k^s}$ before allocation, and processing time $t^{v_k^s}$ of different VNF functions executed in the SFC request $s\in S$. The optimization problem $\mathcal{P}$ is an NP-hard, combinatorial and non-convex problem. To solve this problem of SFC provisioning for large-scale networks in a fast and efficient manner, we propose a scalable distributed design using our previously proposed architecture of DRL model \cite{onsu2024unlocking} that is suitable for use in varying numbers of DCs and logical links of different clusters.

\section{Scalable SFC Provisioning } \label{NovelAI}


\subsection{Distributed Design}
 
To optimize SFC provisioning performance in large-scale networks, a distributed approach is proposed with one general agent and multiple local agents which are explained as follows: 

\subsubsection{General Agent}
The general agent, as the higher-priority agent,  receives overall network information, including all DCs and logical links among them. Initially, it divides the network into a predefined number of clusters based on the DC locations utilizing the Constrained K-Means algorithm \cite{constrainedKMEANS} to ensure balanced cluster sizes. Subsequently, each cluster is assigned a local agent consisting of a DRL module to handle its SFC provisioning tasks.  This completes the setup, with the network segmented into clusters and each cluster managed by a local agent. From this point, the general agent is primarily responsible for observing the system and the actions of the local agents. It mainly assists local agents in path-finding for the packet transmission to a destination DC that is outside the scope of the local agent's cluster. Also, a local agent communicates to the general agent if its resources are not sufficient to handle more SFC requests, then the general agent transfers these requests to other available local agents. The general agent also helps with performance analysis for each local agent by holding their information related to the number of SFCs they satisfy or drop. 

\subsubsection{Local Agents}

Each local agent is responsible for handling the SFC requests generated within the DCs of its clusters as the source address. The destination address of SFC requests can be a DC inside or outside its cluster.  Each local agent is limited to their own DCs and logical links, with no knowledge of neighboring clusters. All local agents operate independently and communicate only with the general agent. A local agent handles the SFC provisioning tasks via proper VNF placement to its DCs utilizing the DRL module and SFC provisioning algorithm provided in our previous work \cite{onsu2024unlocking}. For a quick reference, a brief description of DRL module architecture is provided in subsection \ref{SEC:alg}. Furthermore, if the SFC provisioning task needs assistance beyond the local agent's scope such as packet transmission outside its cluster or handling SFC requests beyond its resource constraints, it communicates with the general agent for path finding or using available resources of other clusters, respectively.

\subsubsection{Path Discovery and Cluster-to-Cluster Packet Transmission}

For packet transmission, selecting the optimum path through logical links, defined as the shortest path with available bandwidth, is crucial. In standard shortest path algorithms, each DC is treated as a node, and each connection as a bidirectional link. To find the shortest path, Dijkstra's algorithm identifies all possible paths and selects the shortest one, we name this process as DC-to-DC (D2D) path discovery. Despite its effectiveness for small-scale networks, its complexity renders it time-consuming for large-scale networks. To address this issue, this research introduces the Cluster-to-Cluster (C2C) path discovery method.  When the source and destination are within the same cluster and the available path exists, the standard method is applied. 
Otherwise, the C2C path discovery method is employed first. In this method, a local agent communicates with a general agent, which determines the path from the cluster of the source DC to the cluster of the destination DC using the Depth First Search (DFS) algorithm. This is followed by iterative traversal from cluster to cluster using D2D path discovery within each cluster until the destination is reached.  Moreover, in iterative traversal from cluster to cluster, only DCs of two clusters are forwarded to the D2D path discovery algorithm, which reduces the scale and complexity of the overall path-finding. 
Thus, the proposed path-finding method incorporates both C2C and D2D path discovery techniques. By focusing on neighboring clusters rather than the entire network, path discovery time is significantly reduced. The complexity of the DFS algorithm for C2C path-finding is \(O(C + L_x)\), where \(C\) is the total number of clusters and \(L_x\) represents the number of inter-cluster links. Within two neighboring clusters, complexity of D2D path discovery is $O((N_{ab}+L_{ab})\times \log N_{ab})$, where \(N_{ab}\) is the total number of DCs and \(L_{ab}\) is the total number of links in the two combined clusters \(a\) and \(b\).

\begin{algorithm}
\caption{SFC Provisioning in Distributed Design}\label{alg:vnf-placement}
\fontsize{7.5}{7.7}\selectfont
\KwData{Overall\_SFCs\_info, Network, C\_Limit}
\KwResult{SFCs requests  $\gets$ accepted or dropped}

 $General\_Agent \gets General\_Agent\_initialization()$\;
 $Clusters\_of\_Network \gets Network.make\_clusters(C\_Limit)$\;
 $General\_Agent.Local\_Agents \gets \{\}$\;
\ForEach{$(id, dataCenters) \in Clusters\_of\_Network$}
{
     $Local\_Agent_{id} \gets Local\_Agent\_Initialization(id, DCs$  
    \\\phantom{forforforforforforforforforforforforfor}$Get\_DRL\_Model())$\\
     $Local\_Agent_{id}.Assigned\_SFC(Cluster\_SFCs\_info, DCs)$\\
    $General\_Agent.Local\_Agent\_Dict[id] \gets Local\_Agent_{id}$\\
}
\While{SFCs\_exists}
{
    \ForEach{$agent \in General\_Agent.Local\_Agents$}
    {
         $(return, act) \gets agent.SFC\_provisioning\_Algo()$\;
        \If{$return = -1$}
        {
             $General\_Agent.perform\_action(agent, act)$\;
        }
    }
     $General\_Agent.collect\_system\_information()$\;
}
\end{algorithm}

\subsection{Distributed SFC Provisioning Algorithm} \label{SEC:alg}
The distributed design algorithm for SFC provisioning can be seen in Algorithm \ref{alg:vnf-placement}. Here, overall SFC information, Network information (all DCs and logical links), and cluster size limit are given as input data. In the beginning, the General Agent is initialized, and the network is split into several clusters considering the cluster size (Line 1-2). Each cluster has its DCs list and cluster ID, and these are assigned alongside the DRL module to initialize the local Agents during the iteration of each cluster (lines 4-6). After initialization, SFC requests from its cluster DCs are assigned to the local agent (line 7), and information of each local agent is added to the General Agent observation module (line 8). Once all local agents are generated, the SFC provisioning algorithm \cite{onsu2024unlocking} starts running for each cluster independently. Each agent adopts the DRL module for satisfying the SFC requests and updating the environment (line 12). However, if the task is beyond the capability of the local agent (return = -1), such as packet transmission beyond the scope of its DCs or resource constraints for further VNF placement, General Agents assist the local agent (line 14) by path-finding and utilizing nearby clusters available resources. At the end of each step, the General Agent gathers environmental data from all local agents (line 17). 

\begin{figure}
    \centering
    \includegraphics[width=0.90\linewidth]{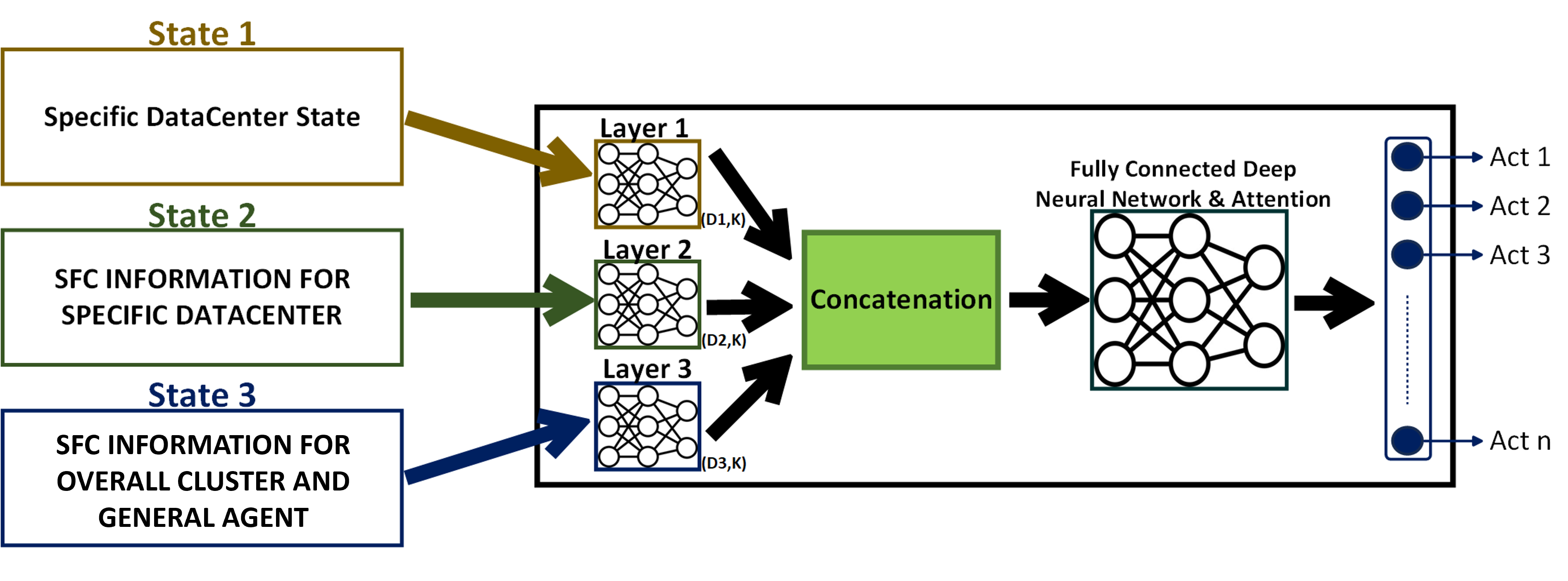}
    \caption{Advanced DRL model architecture with multiple inputs }
    \label{fig: model architecture}
\end{figure}

A concise overview of the DRL module used for SFC provisioning is now provided. For DRL architecture, fully connected deep neural network (FCDNN) layers are used with three input layers and a single output layer. 
First, inputs with different dimensions are forwarded to normalization before being sent to input layers. These layers include the same number of output neurons to provide the same number of output features. Then, the outputs of these layers are concatenated to form one instance forwarded to the attention layer to emphasize the important features. Then it goes through several fully connected DNN hidden layers until reaching the output layer. The DRL model architecture can be seen in \figurename \ref{fig: model architecture}.

Each local agent shares the same DRL architecture and performs their actions only on the DCs of their own cluster. For SFC provisioning, the Deep Q-Network (DQN) algorithm is employed to estimate the optimal action-value function or Q-value. The DQN model takes actions based on the current state, with the agent receiving rewards and Q-values to maximize the action-value function during updates. The DC selection procedure and SFC provisioning algorithm are provided in our previous work \cite{onsu2024unlocking}. A brief explanation of the states, actions, and rewards within the DRL model is provided below.

The state serves as the model’s input layer containing different kinds of system information and these information are challenging to merge. Therefore, multiple input layers of varying sizes are utilized to prevent information loss. In total, there are three different input layers: the first two inputs provide details about the current DC where the algorithm performs actions about its incoming SFC requests, and available resources, while the third input layer contains SFCs information of its cluster DCs and the general agent.  These different kinds of information include E2E latency, bandwidth requirement, VNFs situation for each type in the chain, installed VNFs, and resource availability.
The DRL module’s actions include placing or uninstalling a VNF function from the selected DC, with idle waiting also considered as an action. The number of neurons in the output layer of the DRL corresponds to the number of possible actions and is defined as [$2*|V|+1$], where $|V|$ represents the number of VNF types. The model’s output specifies both action type (placing or uninstalling) and  VNF type. Following this, priority points \cite{onsu2024unlocking} are assigned to all VNFs of the selected type across different SFCs before the action is executed.

A reward is a scalar feedback reflecting the performance of an agent performance for each action, with penalties imposed for suboptimal actions. An optimal action, such as fulfilling an SFC request by successfully placing all its VNFs, earns a positive reward of $+2$. In contrast, dropping an SFC request incurs a penalty of $-1.5$, slightly smaller in magnitude than the reward. Additional penalties are applied for actions such as uninstalling necessary VNFs $(-0.5)$ and selecting invalid actions $(-1)$ which leave the state unchanged keeping the model in the same action until the next step.

\section{Numerical Results} \label{sec:4}
\subsection{Simulation Setup}
The simulation model for the SFC provisioning algorithm is created using Python. The network includes DCs, fiber connections, and SFC requests with specific VNF chains. Each DC has 2 TB storage, 12-120 GHz CPU, and 256 GB RAM, while logical links have 1 Gbps BW. The system is divided into clusters with intra-cluster and inter-cluster logical connections. SFC requests are generated from clients as request bundles (Table \ref{tab:1}) and sent to the network for processing \cite{onsu2024new}.

The simulation generates the six most commonly referenced SFCs: CG, AR, VoIP, VS, MIoT, and Ind 4.0 \cite{2}. These requests have specific bandwidth, E2E delay, request bundle range, and VNF sequences shown in Table \ref{tab:1}. VNF sequences include Network Address Translation (NAT), Firewall (FW), Video Optimization Controller (VOC), Traffic Monitor (TM), WAN Optimizer (WO), and Intrusion Detection Prevention System (IDPS). To satisfy a SFC request, its VNFs must be processed in a proper sequence.  The vCPU, RAM storage, and processing time of each VNF are taken from \cite{2}. 
Quick and effective VNF placement/installation and allocation are essential to meet SFC requests within their E2E delay constraint and improve network performance in terms of acceptance ratio (AccRatio).

\begin{table} 
\centering
\caption{Service Function Chain (SFC) characteristics \cite{2}}\fontsize{6.5}{7.7}\selectfont
\begin{tabular}{|p{1.4cm}|p{1.6cm}|p{1cm}|p{1cm}| p{1 cm}|} 
 \hline
 \textbf{SFC Request}&\textbf{VNF Sequence} &\textbf{Bandwith (Mbps)} &\textbf{E2E delay (msec)} &\textbf{Request Bundle} \\  
 \hline
  Cloud Gaming (CG) & NAT-FW-VOC\break-WO-IDPS  & 4 & 80  & [40-55] \\
  \hline
  Augmented \break Reality (AR) & NAT-FW-TM\break-VOC-IDPS & 100 & 10  & [1-4] \\
  \hline
  VoIP & NAT-FW-TM\break-FW-NAT & 0.064 & 100 & [100-200] \\
  \hline
  Video Streaming (VS) & NAT-FW-TM\break-VOC-IDPS & 4  & 100  & [50-100] \\
  \hline
  MIoT & NAT-FW-IDPS  & [1-50] & 5 & [10-15] \\
  \hline
  Ind 4.0 & NAT-FW & 70 & 8 & [1-4] \\
 \hline
\end{tabular}
\label{tab:1}
\end{table}

\subsection{Training}
The algorithm can be applied to any network configuration, so DRL model training is conducted on a system of 2-4 DCs. The model undergoes 350 updates every 20 episodes. Each episode starts with incoming SFC requests and ends when all requests are processed. SFC requests are randomly generated within the specified bundle size given in Table \ref{tab:1}. SFCs requests are satisfied if all VNFs in their chain are processed in time; otherwise, they are dropped due to resource limitations or E2E delay constraints. Each episode features different SFCs volumes and DC counts to enhance training resilience and diversity. Each agent model performs 100 actions per step, with each step lasting 1 ms and an action inference timestamp of 0.01 ms. At the end of each episode, the model records the state, action, next state, and reward observation set to replay memory. During updates, it optimizes parameters for higher future rewards, using random batches from replay memory for training.

\begin{figure}
    \centering
    \includegraphics[width=.85\linewidth]{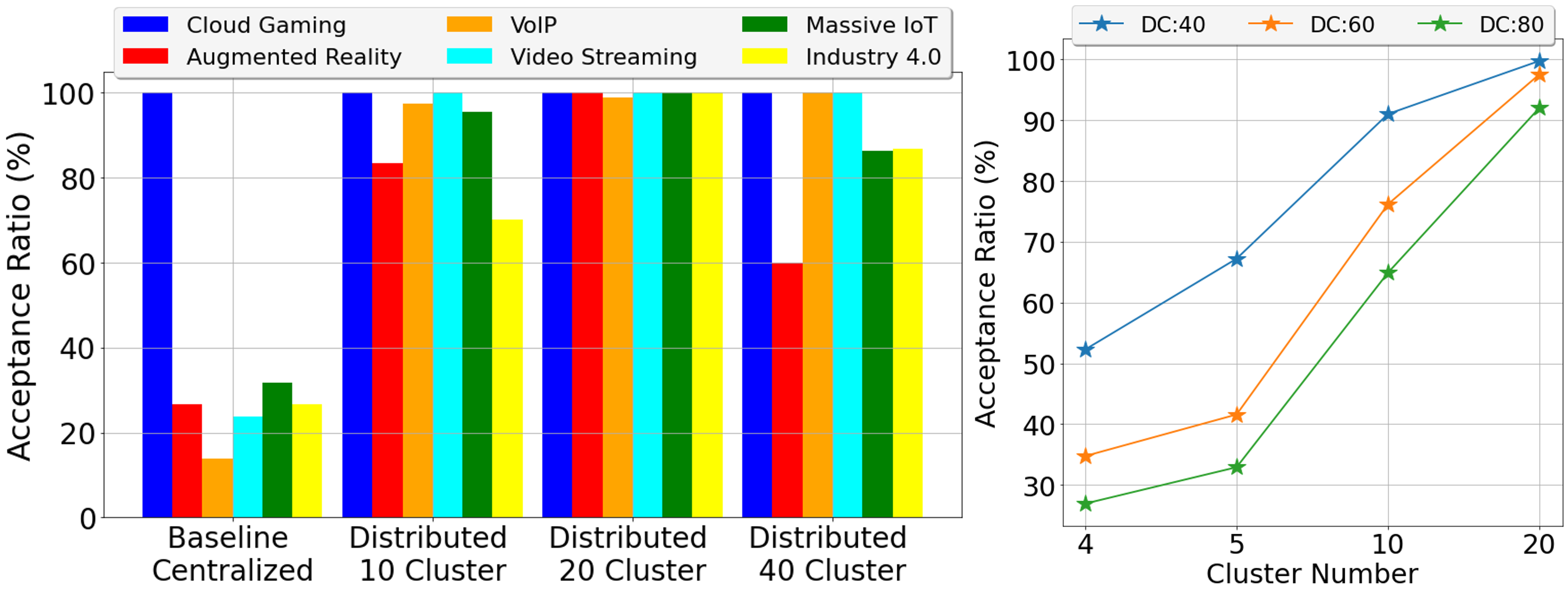}
    \caption{Left: SFC AccRatio comparison to centralized approach for 40 DCs network; Right: SFC AccRatio for varying clusters and  DCs numbers for Distributed Design.}
    \label{fig: res 1}
    \vspace{-0.1in}
\end{figure}

\subsection{Results}
To measure the performance of the proposed approach the following metrics have been utilized: E2E delay, overall SFC AccRatio, and  AccRatio of each SFC type. 
In \figurename \ref{fig: res 1}, the left diagram compares the performance of the proposed distributed design with the baseline centralized approach in terms of AccRatio for each type of SFC  with varying cluster numbers for a network of 40 DCs.  Although the comparison is percentage-based, each SFC type has a different bundle size, as outlined in Table \ref{tab:1}. SFC requests are generated twice, exceeding standard bundle sizes for a more demanding test. In a large-scale setup with 40 DCs, the centralized approach handles only 35.03\% of SFC requests due to computational limits. In contrast, the distributed design achieves 91.04\% with 10 clusters and 99.79\% with 20 clusters. Performance drops to 87.99\% with 40 clusters due to increased communication demands and resource constraints on local agents. Note that in a network with a fixed number of DCs, increasing the cluster numbers is equivalent to decreasing the cluster size i.e. number of DCs per cluster. 

The right diagram in \figurename \ref{fig: res 1} tests the proposed distributed design across 40, 60, and 80 DCs networks with varying total cluster numbers such as 4, 5, 10, and 20. As the network enlarges, SFC requests increase: 1.5 times for 60 DCs and 2 times for 80 DCs. Networks with 4 and 5 clusters show low AccRatio  (below 40\%) for 60 and 80 DCs due to the large number of DCs in each cluster. It can be visualized that with an increase in the number of clusters from 10 to 20 clusters, AccRatio improves up to above 90\% which justifies the use of the proposed distributed design in a large-scale network. 

\begin{figure}
    \centering
    \includegraphics[width=0.97\linewidth]{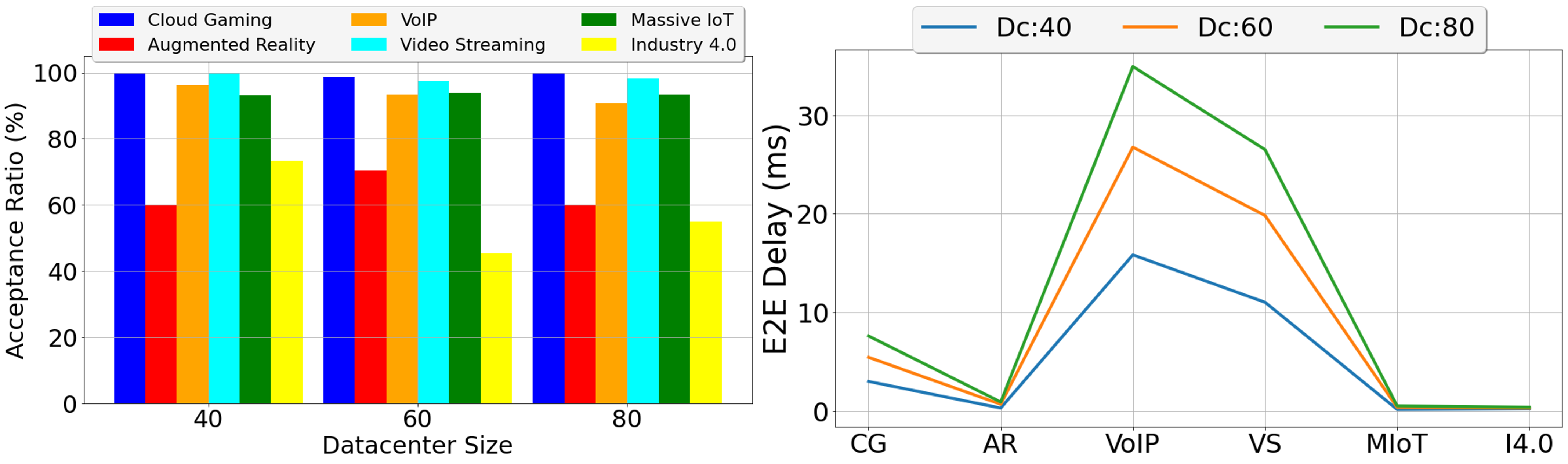}
    \caption{SFC AccRatio and E2E delay under different DCs number }
    \label{fig: res 2}
\end{figure}

In \figurename \ref{fig: res 2}, SFC AccRatio and E2E delay are presented for different DC numbers with a fixed cluster size of 5 DCs, meaning each local agent manages 5 DCs.  As the network scales in terms of total DCs, the number of clusters/local agents also increases. The left diagram shows AccRatio of each SFC type, where CG has the highest AccRatio due to its 80 ms delay tolerance which is lower than that of VS and VoIP (100 ms) making low E2E delay tolerant requests prioritized. AR and Ind 4.0 have the highest drop ratios due to strict E2E delay limits. MIoT, with minimal delay, is processed earlier, resulting in a higher AccRatio than AR and Ind 4.0. In the right diagram, the E2E delay of accepted SFCs is presented. The E2E delay increases with more DCs due to higher request volumes and transmission times. VoIP and VS show higher E2E delay due to their delay attributes, causing later processing.

\begin{figure}
    \centering
    \includegraphics[width=0.97\linewidth]{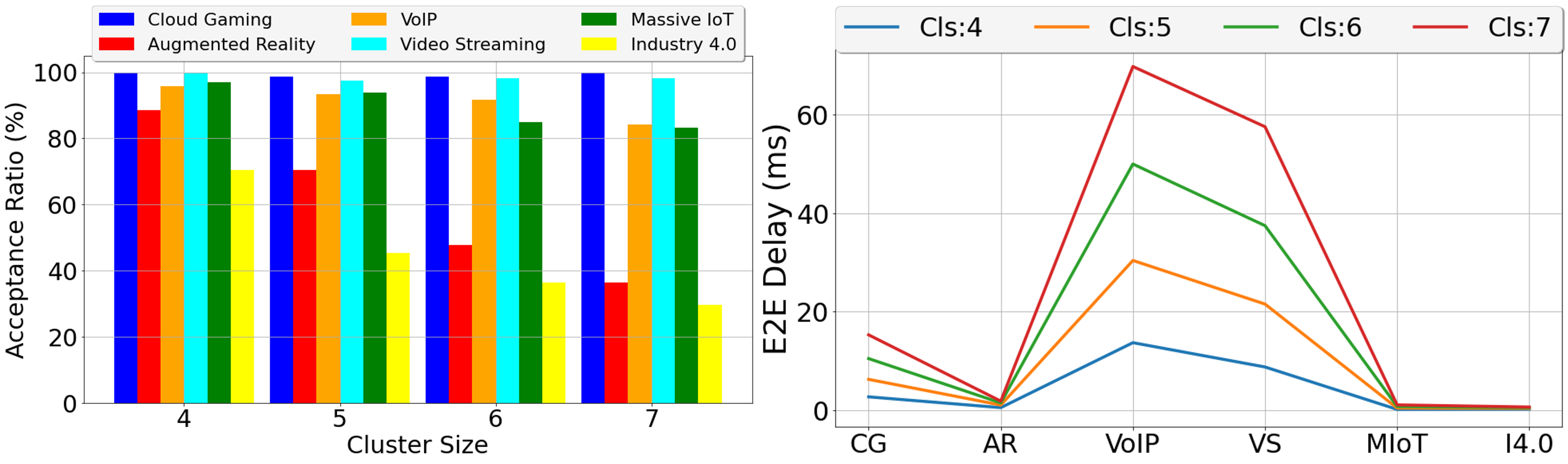}
    \caption{SFC AccRatio and E2E delay under different cluster sizes for local agents}
    \label{fig: res 3}
\end{figure}
 
The effect of different cluster sizes on performance metrics is demonstrated in \figurename \ref{fig: res 3}. A larger cluster size assigns more DCs to a local agent and reduces the number of local agents, making the system resemble a centralized approach. This reduces communication with general agents but increases demands for one local agent to handle, leading to fewer accepted SFCs, particularly for low E2E delay requests such as Ind 4.0 and AR. Additionally, E2E delays increase with larger cluster sizes.

\section{Conclusions} \label{sec:5}

This paper has explored scalable SFC provisioning in large-scale networks using a DRL model and priority-based VNF placement to handle many SFC requests. A custom simulation with data centers, connections, and SFC requests was created, training the model on SFC attributes and resource constraints. To ensure scalability, tasks were divided among local agents supervised by a general agent, improving performance. Results showed larger networks slow down the model and hinder SFC provisioning, but task division and effective communication enhance efficiency. Our ongoing work focuses on the resource constraints of local agents and dynamic cluster sizes.

\section*{Acknowledgment }\label{Section6}
This work is supported in part by the Natural Sciences and Engineering Research Council of Canada (NSERC) Alliance Program, in part by the   MITACS Accelerate Program, and in part by the NSERC CREATE TRAVERSAL program.


\bibliographystyle{IEEEtran}


\end{document}